\documentclass[conference]{IEEEtran}
\IEEEoverridecommandlockouts

\usepackage{cite}
\usepackage{amsmath,amssymb,amsfonts}
\usepackage{graphicx}
\usepackage{booktabs}

\usepackage{algorithm}
\usepackage{algpseudocode}

\usepackage{textcomp}
\usepackage{xcolor}
\usepackage[hidelinks]{hyperref}
\usepackage{makecell}

\def\BibTeX{{\rm B\kern-.05em{\sc i\kern-.025em b}\kern-.08em
		T\kern-.1667em\lower.7ex\hbox{E}\kern-.125emX}}
\begin{document}

\title{Pre-Filtering Code Suggestions using Developer Behavioral Telemetry to Optimize LLM-Assisted Programming}

\author{\IEEEauthorblockN{Mohammad Nour Al Awad}
	\IEEEauthorblockA{
		\textit{ITMO University}\\
		Saint Petersburg, Russia \\
		MohammadNourAlAwad@itmo.ru}
	\and
	\IEEEauthorblockN{Sergey Ivanov}
	\IEEEauthorblockA{
		\textit{ITMO University}\\
		Saint Petersburg, Russia \\
		svivanov@itmo.ru}
	\and
	\IEEEauthorblockN{Olga Tikhonova}
	\IEEEauthorblockA{
		\textit{ITMO University}\\
		Saint Petersburg, Russia \\
		tikhonova\_ob@itmo.ru}
}
\maketitle
\begingroup
\renewcommand\thefootnote{}\footnote{
	© 2025 IEEE. Personal use of this material is permitted.
	Permission from IEEE must be obtained for all other uses.
}\addtocounter{footnote}{-1}
\endgroup

\begin{abstract}
Large Language Models (LLMs) are increasingly integrated into code editors to provide AI-powered code suggestions. Yet many of these suggestions are ignored, resulting in wasted computation, increased latency, and unnecessary interruptions. We introduce a lightweight pre-filtering model that predicts the likelihood of suggestion acceptance before invoking the LLM, using only real-time developer telemetry such as typing speed, file navigation, and editing activity. Deployed in a production-grade Visual Studio Code plugin over four months of naturalistic use, our approach nearly doubled acceptance rates (18.4\% → 34.2\%) while suppressing 35\% of low-value LLM calls. These findings demonstrate that behavioral signals alone can meaningfully improve both user experience and system efficiency in LLM-assisted programming, highlighting the value of timing-aware, privacy-preserving adaptation mechanisms. The filter operates solely on pre-invocation editor telemetry and never inspects code or prompts.
\end{abstract}


\section{Introduction}

Large Language Models (LLMs) have rapidly transformed the landscape of software development by enabling intelligent code completions, refactorings, and in-editor conversations. These capabilities are increasingly integrated into modern development environments, particularly through plugins for popular IDEs such as Visual Studio Code. However, despite their power, LLM-driven code suggestions often fail to align with developer intent in real-time, leading to low acceptance rates, disrupted workflows, and wasted computational resources \cite{ziegler2024copilotproductivity}.

Recent advances in LLM optimization have focused on improving model quality, response latency, or retrieval augmentation. Yet, much less attention has been given to the human side of the interaction—specifically, whether the developer is cognitively ready, task-wise situated, or contextually interested in receiving a suggestion at a given moment. Treating all editing contexts as equally appropriate for suggestion triggering is both computationally expensive and interactionally naive \cite{mozzannar2024whentoshow}. Large-scale enterprise deployments, such as those studied by Weisz \emph{et al.} \cite{weisz2025examininguseimpactai}, confirm that usage context and developer workflows strongly shape the effectiveness and perceived value of AI-assisted coding tools.

In this paper, we propose a lightweight behavioral pre-filtering mechanism designed to operate prior to LLM invocation. Instead of analyzing prompt content or generated code, our system uses real-time behavioral telemetry to predict whether a developer is likely to accept a suggestion. This prediction serves as a gatekeeper: if the model forecasts low likelihood of acceptance, the suggestion request is suppressed, thus saving LLM resources and reducing interface interruptions.

Our approach is grounded in a human-centered view of developer-AI interaction. We model the developer not as a passive recipient of completions, but as an active agent engaged in cognitively demanding tasks—navigating files, writing code, pausing to think, reacting to errors, and using IDE features. These interaction patterns are observable and measurable through standard APIs in the development environment, allowing us to capture a continuous representation of the developer's state without manual labeling or intrusive interventions.

To evaluate our approach, we deployed the pre-filtering model inside a production-grade VS Code plugin used across our internal research center. The model was trained on months of sparsely collected telemetry data and suggestion logs. We compare usage before and after deployment of our trained model, revealing a significant improvement in suggestion acceptance rates—from 18.4\% to 34.2\%—and a reduction of approximately 35\% in LLM suggestion calls. These results suggest that even in noisy and non-scripted environments, behavioral context alone can meaningfully improve the efficiency and usability of AI-powered programming tools.

Our contributions are as follows:
\begin{itemize}
	\item We introduce a behavior-first framework for pre-filtering code suggestions, operating entirely without prompt or model content.
	\item We develop and deploy a production-ready system that captures, processes, and aligns behavioral telemetry in real-world VS Code usage.
	\item We demonstrate substantial improvements in acceptance rates and resource efficiency, showing that behavioral signals can serve as reliable proxies for user readiness and intent.
\end{itemize}

This work opens new directions for modeling human-AI interaction in coding environments, emphasizing adaptivity, efficiency, and respect for developer context as first-class design goals.

The remainder of this paper is organized as follows: Section~\ref{sec:related} reviews related work on adaptive suggestion systems and developer modeling. Section~\ref{sec:behavioral} outlines our telemetry design, behavioral framing, and modeling rationale. Section~\ref{sec:deployment} describes the deployment methodology and evaluation setup. Section~\ref{sec:results} presents our findings and discusses the implications. Section~\ref{sec:conclusion} concludes with reflections and future directions.

\section{Related Work}
\label{sec:related}

\subsection{AI-Powered Code Suggestions and Developer Acceptance}

LLM-based coding assistants such as GitHub Copilot, Amazon CodeWhisperer, and Cursor have reshaped the way developers interact with code editors, enabling real-time suggestions for code completion, refactoring, and documentation. Despite their growing popularity, multiple empirical studies report relatively modest acceptance rates\cite{yetistiren2023codequality} in real-world usage. For example, field deployment at ZoomInfo found that only around one-third of Copilot suggestions were accepted by developers during daily work \cite{bakal2025copilotzoominfo, barke2023grounded}. Broader ecosystem analyses, such as Song \emph{et al.}’s study of GitHub Copilot adoption in open-source projects \cite{song2024oss}, further emphasize that the impact of LLM-based assistance is shaped by collaborative and project-level factors. Recent work has also combined prompt-based models with multi-retrieval augmentation to increase completion accuracy and contextual relevance \cite{10.1145/3725812}. Other work has shown that the way suggestions are presented influences acceptance: Oertel et al. demonstrated that encouraging developers to explore more than the first candidate improves the quality of adopted completions without harming productivity \cite{oertel2025dontsettle}.

While these studies focus on optimizing presentation or tuning suggestion ranking, our work takes a complementary direction by addressing the triggering decision itself—predicting whether any suggestion should be generated in the first place, based solely on real-time behavioral signals.

\subsection{Telemetry-Driven Observability in IDEs}

The increasing integration of AI tools into developer environments has sparked efforts to incorporate observability and telemetry as first-class entities. Koc et al. introduced the Model Context Protocol (MCP) \cite{koc2025mcp}, advocating for embedding metrics and runtime context into LLM-enhanced IDEs to better understand prompt behavior and system load \cite{koc2025mcp}. Nam et al. analyzed fine-grained usage data from Google’s code editor, revealing struggles with prompt-driven editing and proposing corrective workflows \cite{nam2025promptediting}. These works underscore the value of instrumentation in LLM systems, yet typically treat telemetry as a post-hoc debugging signal. In contrast, we use behavioral telemetry as a real-time signal to modulate LLM invocation itself.

\subsection{Modeling Developer Behavior}

A long-standing thread in software engineering has studied developer interactions to model cognitive states and predict future behavior. Earlier works used IDE command logs to predict developer intent \cite{bulmer2020idecommands}, or topic models to infer upcoming tasks from code navigation patterns \cite{damevski2017topicide, meyer2017worklife}. Others have connected temporal features such as typing bursts or navigation frequency to self-reported productivity \cite{meyer2017worklife}. Similarly, Overwatch \cite{10.1145/3563302} leverages patterns in sequential code edits to predict future actions, underscoring the predictive power of temporal interaction traces.
Our approach builds on this tradition by using observable developer activity—typing, pauses, corrections, and navigation—as a real-time representation of engagement and task context, to predict suggestion receptiveness. However, unlike prior work focused on retrospective prediction, our pre-filtering method operates online and is embedded within a deployed LLM-powered assistant.

\subsection{Interruptibility and Context-Aware Assistance}

A parallel line of research in human-computer interaction explores the timing of assistance and its impact on cognitive load. Fogarty et al. pioneered the use of sensors to infer human interruptibility and modulate desktop notifications accordingly \cite{fogarty2005interruptibility}. Subsequent work confirmed that interruptions during software development—especially those unrelated to the immediate task—can fragment attention and reduce perceived productivity \cite{meyer2017worklife}. Also misalignment between a developer’s mental model of the AI assistant and its actual behavior can reduce trust and uptake; elicitation studies by Desolda \emph{et al.} \cite{2025arXiv250202194D} highlight these cognitive factors as critical in designing timing-aware suggestion systems. 
Our system can be interpreted as a software-level analog to these models: it withholds suggestions during behavioral patterns suggestive of low interruptibility (e.g., high error load, frequent corrections, or rapid typing), thereby aligning system behavior with developer readiness.

\subsection{Positioning and Novelty}
In summary, prior work has emphasized three key themes: (i) measuring and improving LLM suggestion acceptance, (ii) instrumenting developer environments with observability and telemetry, and (iii) modeling developer behavior using IDE logs. This work bridges these strands by introducing a proactive behavioral gatekeeper that uses real-time telemetry to determine whether a suggestion should be shown. To our knowledge, it is the first deployed system to use solely developer behavioral state—rather than prompt content or model confidence—as the basis for suppressing or triggering LLM completions. Recent work by de Moor \emph{et al.} has explored a complementary approach, using a transformer-based model to predict the optimal moments for invoking code completion based on IDE telemetry, highlighting the value of timing-aware suggestion systems \cite{demoor2024smart}. Our findings show that such a filter can substantially reduce computational overhead while improving the alignment between suggestions and developer intent.

\section{Behavioral Metrics and Modeling}
\label{sec:behavioral}

\subsection{System Architecture}
\label{sec:behavioral-architecture}

Our pre-filtering model is integrated into a production-grade, LLM-powered coding assistant delivered via a Visual Studio Code plugin. The system operates in real time and comprises several modular components. The plugin continuously captures behavioral telemetry from the developer's interaction with the IDE, applies heuristic constraints (e.g., max token length), and uses an adaptive mechanism to modulate suggestion triggering timing.

To this architecture, we added a lightweight behavioral pre-filtering model that predicts whether the developer is likely to accept a suggestion—before invoking the LLM. If the model forecasts low acceptance likelihood, the suggestion request is suppressed entirely, thereby reducing unnecessary GPU usage and latency. If the request passes the filter, it proceeds to a generation pipeline that includes a retrieval-augmented generation (RAG) component and Qwen LLMs served via vLLM for low-latency inference. Post-generation, suggestions are optionally trimmed or reformatted to improve UX.

While the full assistant includes additional modules—such as personalized code indexing, dynamic suggestion shaping, and real-time feedback logging—this paper focuses exclusively on the behavioral pre-filtering component and its impact on system efficiency and suggestion quality.

\subsection{Data Collection Procedure}

To understand the behavioral context surrounding code suggestion events, we instrumented a Visual Studio Code extension to log two primary types of telemetry: (i) event-level metrics associated with each suggestion request, and (ii) continuous behavioral summaries aggregated at a one-minute resolution. All data collection was conducted anonymously and in compliance with institutional consent procedures, with no raw code or prompt content retained.

\paragraph{Per-Suggestion Logging.} For every code suggestion request made through the plugin, we captured interaction-level metadata such as the length of the prompt, number of characters in the generated suggestion, time to accept or reject, and whether the suggestion was eventually accepted. A suggestion was marked as \textit{accepted} if the developer inserted it into their code without requesting another suggestion. Conversely, if the developer requested a new suggestion without accepting the previous one, we labeled it as a \textit{rejection} with passive rejection after 30 s of inactivity. This framing aligns with natural developer behavior in IDEs, where the act of bypassing a suggestion implies contextual irrelevance or dissatisfaction.

\paragraph{Continuous Behavioral Metrics.} In parallel, we recorded a rolling set of behavioral features aggregated every minute, providing a snapshot of the developer’s activity independent of suggestion events. These include typing speed, frequency of pauses, file navigation events, command palette usage (e.g., copy, paste, quick fix), and code compilation outcomes such as warnings or errors. This behavioral stream forms the temporal context in which each suggestion request occurs and serves as the primary input to our pre-filtering model.

\paragraph{Estimating Task Complexity.} To contextualize developer effort, we compute a per-file scalar \emph{Task Complexity} at the time of each autocomplete request. Using Tree-sitter, the surrounding code is parsed and summarized into a single score that combines cyclomatic paths, Halstead effort (operator/operand counts), and a maintainability index. The method supports all our supported languages, and falls back to heuristics based on decision keywords and LOC when a grammar is unavailable. Only aggregate metrics are retained—no code or identifiers—making the feature language-aware but content-agnostic.

\subsection{Behavioral Categories and Feature Design}
\label{sec:behavioral-categories}

Rather than relying on code content, prompt formulation, or model internals, our system is driven by real-time telemetry reflecting how developers engage with the IDE environment. The classifier consumes a rich stream of behavioral signals captured prior to suggestion triggering—offering a proactive, privacy-preserving proxy for developer receptiveness.

We group the engineered features into five conceptual categories that reflect different dimensions of interaction: fluency (e.g., typing rhythm), editing scope, command usage, code state, and session context. Table~\ref{tab:behavioral_features} summarizes representative features across these categories.

Together, these metrics form a compact yet expressive representation of developer behavior. Since all features are computed before LLM invocation, the model operates in real time without inspecting prompt or code content, i.e. all predictors are purely behavioral and were chosen for interpretability and cross-language applicability.

\begin{table}[ht]
	\centering
	\caption{Behavioral categories and representative features used by the acceptance classifier}
	\label{tab:behavioral_features}
	\begin{tabular}{@{}p{0.12\textwidth} p{0.12\textwidth} l@{}}
		\toprule
		\textbf{Category} & \textbf{Purpose} & \textbf{Example Features} \\
		\midrule
		Interaction Fluency & 
		Typing pace and continuity & 
		\makecell[l]{Typing speed,\\ Total characters typed,\\ Pause count,\\ Typing efficiency} \\
		
		Code Editing and Scope & 
		Engagement with code structure and files & 
		\makecell[l]{Lines of code added,\\ File size,\\ Edit density,\\ Number of open files} \\
		
		IDE Command Usage & 
		Use of assistive and tooling features & 
		\makecell[l]{Undo frequency,\\ Quick fix usage,\\ Terminal toggling,\\ Command palette actions} \\
		
		Code State & 
		Stability and complexity of current code & 
		\makecell[l]{Number of warnings/errors,\\ Breakpoint count,\\ Estimated code complexity} \\
		
		Session Context & 
		Cumulative indicators across the session & 
		\makecell[l]{Suggestions accepted/rejected,\\ Acceptance ratio,\\ Total typing duration} \\
		
		\bottomrule
	\end{tabular}
\end{table}

\subsubsection*{Engineered Behavioral Ratios}

To capture dynamic aspects of developer interaction, we derive several engineered ratios from raw telemetry. These metrics normalize key behavioral signals relative to contextual baselines (e.g., time, file size), enabling consistent modeling across users and sessions.

\begin{align*}
	\text{Typing Efficiency} &= \frac{C_{\text{typed}}}{T_{\text{typing}} + \epsilon} \\
	\text{Pause Frequency}   &= \frac{N_{\text{pauses}}}{T_{\text{typing}} + \epsilon} \\
	\text{Acceptance Ratio}  &= \frac{N_{\text{accepted}}}{N_{\text{accepted}} + N_{\text{rejected}} + \epsilon} \\
	\text{Edit Density}      &= \frac{L_{\text{added}}}{L_{\text{file}} + \epsilon}
\end{align*}

\noindent
Where:
\begin{itemize}
	\item \( C_{\text{typed}} \): Total characters typed during the aggregation window.
	\item \( T_{\text{typing}} \): Total typing duration in seconds.
	\item \( N_{\text{pauses}} \): Number of pauses exceeding a threshold (e.g., 2s).
	\item \( N_{\text{accepted}} \), \( N_{\text{rejected}} \): Count of accepted and rejected suggestions in the current session.
	\item \( L_{\text{added}} \): Number of lines of code added during the window.
	\item \( L_{\text{file}} \): Total number of lines in the current active file.
	\item \( \epsilon \): A small constant (e.g., \(10^{-6}\)) to prevent division by zero.
\end{itemize}

\subsection{Model Training}
To operationalize the pre-filtering mechanism, we trained a lightweight binary classifier that predicts whether a suggestion is likely to be accepted by the developer based solely on real-time behavioral features. The dataset used for training comprised 2,318 unique suggestion events, collected from user sessions with informed consent. Out of these, 426 suggestions were accepted by developers, while the remaining 1,892 were either explicitly rejected or bypassed by requesting a new suggestion. This significant class imbalance (approximately 1:4.4) motivated the use of balanced learning techniques.

To ensure robust evaluation and threshold selection, the dataset was split into training, validation, and test sets using stratified sampling. Specifically, 64\% of the data was used for training, 16\% for validation, and 20\% for final testing. Models were trained using randomized hyperparameter search over several lightweight classifiers, including XGBoost, LightGBM, CatBoost, and Balanced Random Forest. While all models achieved comparable performance, the CatBoost model was selected for deployment due to its higher trade-off between precision and recall at low thresholds, along with very small model size, which aligns with our goal of confidently suppressing unproductive suggestions.

Given a training dataset \( \{(x_i, y_i)\}_{i=1}^n \), where each \( x_i \in \mathbb{R}^d \) is a behavioral feature vector and \( y_i \in \{0, 1\} \) denotes whether the suggestion was accepted, we train the classifier by minimizing a weighted binary cross-entropy loss:

\begin{equation*}
	\mathcal{L}(\theta) = - \sum_{i=1}^{n} w_{y_i} \cdot \left[ y_i \log f_\theta(x_i) + (1 - y_i) \log (1 - f_\theta(x_i)) \right]
	\label{eq:loss}
\end{equation*}

Here, \( f_\theta(x_i) \in [0, 1] \) is the model’s predicted probability of acceptance, and \( w_{y_i} \) is a class-specific weight used to address the class imbalance. This formulation allows the model to prioritize recall on accepted suggestions while reducing the frequency of low-value completions.

Importantly, we did not seek to optimize for overall accuracy or F1-score. Rather, our primary objective was to reduce the number of low-value suggestion requests—true negatives—without significantly harming the number of accepted suggestions (true positives). 

\subsection{Justification for Behavioral-Only Filtering}

By excluding prompt content, suggestion content, and developer intent labels from our model inputs, we adopt a deliberately minimalist and privacy-preserving design. This decision was informed by several practical and theoretical considerations:

\begin{itemize}
\item Real-time Operability: Behavioral signals are available before the suggestion is generated, making them ideal for proactive gating.

\item Generalizability: Since behavioral patterns are modality-independent, the approach can generalize across languages and codebases without fine-tuning for domain-specific syntax. 

\item Privacy and Transparency: Avoiding the use of raw code or prompts mitigates concerns around data sensitivity and developer surveillance, aligning the system with ethical guidelines for responsible telemetry.

\item Model Simplicity and Interpretability: Behavioral features are inherently interpretable (e.g., typing speed or undo frequency), allowing system designers to reason about model behavior and thresholds.
\end{itemize}

\section{Experimental Design and Deployment}
\label{sec:deployment}

\subsection{Participants}

Our adaptive suggestion system was deployed in a naturalistic academic research setting. Rather than recruiting a fixed participant group for a scripted study, we made the tool available to a pool of 25 researchers and developers via our internal communication channels. Lab members could freely install and use the plugin during their daily coding tasks. Usage was entirely voluntary and varied organically: some developers tried the system once or twice, while others used it regularly for extended periods.

Prior to the deployment, we surveyed 44 developers across multiple university research environments to identify those who regularly use Visual Studio Code as their primary IDE. From these, 25 were invited to participate and provided with instructions and technical support. Ultimately, 9 experienced developers actively used the assistant for a sufficient period, each contributing at least 50 recorded code suggestions. All active participants had at least 3 years of programming experience (with 78\% reporting over 5 years) and coded for more than 4 hours daily in most cases. Python was the dominant language among them, complemented by JavaScript, TypeScript, C++, and C. Notably, while all participants had prior experience with AI chat assistants (e.g., ChatGPT), most had not used AI-powered IDE plugins before, highlighting the novelty and practical relevance of the adaptive timing mechanism in real workflows.

All usage logs were anonymized and collected with informed consent in accordance with institutional policies.

\subsection{Metric Suite and Threshold Selection}
\label{sec:metrics}

While standard classification metrics provide insight into model discrimination, interactive evaluation frameworks such as the RealHumanEval benchmark \cite{anonymous2024the} demonstrate the importance of measuring assistance quality in realistic, human-in-the-loop coding scenarios. We evaluate our model in alignment with its primary role as a gatekeeper—deciding \emph{whether} to trigger a suggestion—we adopt a metric suite that emphasizes both \emph{ranking quality} and \emph{class-specific utility} under imbalance, rather than global accuracy.

Specifically, we report:

\begin{itemize}
	\item \emph{ROC--AUC}: Measures general discriminative ability under balanced priors.
	\item \emph{PR--AUC}: Captures precision-recall trade-offs under real class imbalance.
	\item \emph{Balanced Accuracy}, \emph{Matthews Correlation Coefficient (MCC)}, and \emph{Cohen’s $\kappa$}: Assess calibration against random or uninformative classifiers.
	\item \emph{Brier Score}: Evaluates the quality of predicted probabilities.
\end{itemize}

Given the asymmetric costs of false positives and false negatives, we selected a deliberately low operating threshold ($\tau = 0.1$) on the validation set. This threshold maximizes recall for \texttt{accepted} suggestions (to avoid suppressing useful completions) while preserving high precision for the more frequent \texttt{rejected} class.

The decision rule for triggering the LLM can thus be formalized as:
\[
\text{Trigger LLM} \iff \hat{P}_{\text{accept}}(x_{\text{behavioral}}) > \tau
\]
where \( \hat{P}_{\text{accept}} \) is the predicted acceptance probability based solely on behavioral features \( x_{\text{behavioral}} \).

\section{Results and Discussion}
\label{sec:results}

We evaluate the impact of our pre-filtering system using two complementary approaches:
\textit{(i)} offline analysis of model performance on a held-out test set, and
\textit{(ii)} live deployment analysis in real developer workflows.

\subsection{Offline analysis of Classifier performance}

\subsubsection{Evaluation Metrics}
\label{sec:metrics_eval}

Evaluation on a held-out test set (464 suggestions) shows:

\begin{table}[ht]
	\centering
	\caption{Classifier Evaluation Metrics on Held-out Test Set}
	\label{tab:metrics}
	\begin{tabular}{lcc}
		\toprule
		\textbf{Metric} & \textbf{Value} & \textbf{Std. Dev.} \\
		\midrule
		ROC-AUC & 0.726 & $\pm$ 0.05 \\
		PR-AUC & 0.350 & $\pm$ 0.09 \\
		Balanced Accuracy & 0.691 & -- \\
		MCC & 0.310 & -- \\
		Cohen $\kappa$ & 0.191 & -- \\
		Brier Score & 0.139 & -- \\
		\bottomrule
	\end{tabular}
\end{table}

These results confirm that despite moderate headline scores, the model provides valuable \textit{ranking utility} under class imbalance.

\subsubsection{Threshold Selection and Gatekeeping Behavior}

Operating at a tuned threshold of $\tau = 0.10$:

\begin{itemize}
	\item \textbf{True Negative Rate (Rejected Class)}: 41.6\% of all rejections filtered before LLM invocation (precision = 0.981)
	\item \textbf{False Negative Rate (Accepted Class)}: Only 3.5\% of accepted suggestions mistakenly filtered (recall = 0.965)
\end{itemize}

This asymmetry is desirable: filtering unproductive suggestions is cheap, while suppressing helpful ones risks user trust.

\subsection{Understanding Behavioral Drivers}
\label{sec:drivers}

To better understand which behavioral factors contributed most to the model’s decision-making, we conducted two complementary analyses of feature influence: permutation-based importance and SHAP (SHapley Additive exPlanations) value interpretation. Both techniques consistently identified several developer behaviors as key predictors of suggestion receptiveness.

\begin{figure}[ht]
	\centering
	\includegraphics[width=0.48\textwidth]{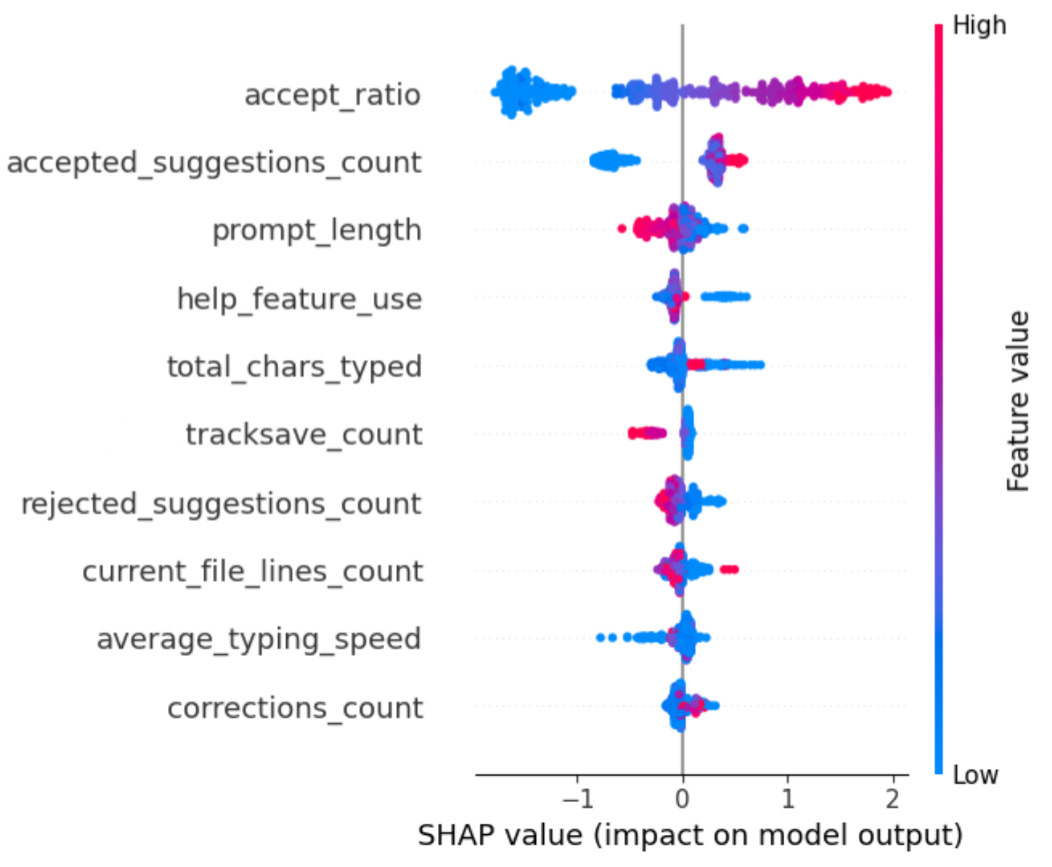}
	\caption{SHAP value analysis for the CatBoost acceptance classifier.}
	\label{fig:features}
\end{figure}

As shown in Figure~\ref{fig:features}, recent acceptance ratio and suggestion history emerged as dominant predictors, with typing rhythm and help-feature usage playing secondary roles. We expand on these behavioral drivers below.

\paragraph{Recent Acceptance Trends.}
The most influential predictor was the developer’s recent history of accepting or rejecting suggestions. A high frequency of accepted suggestions typically indicated continued openness to assistance, whereas a streak of rejections often signaled disengagement. This highlights a momentum effect in developer behavior: prior acceptance is a strong cue for future receptiveness.

\begin{equation*}
	\text{Acceptance Ratio} = \frac{N_{\text{accepted}}}{N_{\text{accepted}} + N_{\text{rejected}} + \epsilon}
	\label{eq:accept_ratio}
\end{equation*}

\paragraph{IDE Tool Usage and Problem-Solving Signals.}
Frequent engagement with built-in help functions or automated quick fixes was associated with a lower likelihood of accepting suggestions. These behaviors may reflect task difficulty, uncertainty, or cognitive load—contexts in which developers are less receptive to interruptions or unsolicited input. This finding resonates with Tang \emph{et al.} \cite{10714560}, who observed that developers engage in extensive validation and repair activities when presented with AI-generated code, especially during periods of high problem-solving activity.

\paragraph{Typing Rhythm and Flow.}
Typing speed and the presence or absence of pauses provided insight into the developer’s cognitive state. Sustained high-speed typing with minimal pauses suggested focus or flow, a state in which the developer is less likely to welcome external suggestions. In contrast, slower or fragmented typing often coincided with a higher likelihood of suggestion acceptance.

\paragraph{Cumulative Activity Context.}
Longer-term interaction patterns—such as the overall volume of code written or the cumulative number of previously accepted completions—also proved informative. These signals helped the model contextualize short-term decisions within the broader scope of a coding session.

Taken together, these results validate the central premise of our approach: that real-time developer behavior offers rich, interpretable signals for deciding when AI assistance is likely to be welcome. The model’s reliance on recent interactions, typing dynamics, and help-seeking patterns reflects a nuanced behavioral framing of suggestion timing—one that respects user attention and reduces unnecessary cognitive load.

\subsection{Live Deployment Analysis}
\subsubsection{Improved Suggestion Acceptance and Call Reduction}
\label{sec:acceptancerate}

After deployment of the pre-filtering mechanism, we observed two complementary gains: a higher acceptance rate on the suggestions that were issued, and a substantial drop in total LLM invocation.

First, of the 2,319 suggestion events recorded in the ``before'' period, all resulted in an LLM call and yielded an 18.4\% acceptance rate. During the ``after'' period, the filter flagged 768 out of 2,190 incoming suggestion requests (35.1\%), suppressing their upstream LLM calls. The remaining 1,422 requests proceeded to the LLM, and of those the developers accepted 34.2\%. Thus:

\begin{itemize}
	\item \textbf{Absolute acceptance increase:} +15.8\,pp (18.4\% \textrightarrow\ 34.2\%)
	\item \textbf{Relative acceptance improvement:} +86\%
	\item \textbf{LLM-call reduction:} 768 calls suppressed (35.1\% of requests)
\end{itemize}

\begin{table}[ht]
	\centering
	\caption{Comparison of suggestion delivery and acceptance before vs.\ after filtering}
	\label{tab:filtering_summary}
	\begin{tabular}{lccc}
		\toprule
		\textbf{Metric} & \textbf{Before} & \textbf{After} \\
		\midrule
		Total suggestion events logged & 2,319 & 2,190\\
		LLM calls actually issued & 2,319 & 1,422\\
		Filtered out suggestions & 0 & 768 \\
		Acceptance rate & 18.4\% & 34.2\% \\
		\bottomrule
	\end{tabular}
\end{table}

By suppressing 768 low-value requests, we reduced GPU usage, network traffic, and backend queuing by nearly 35\%, while simultaneously nearly doubling the yield of accepted completions per LLM call. This dual benefit—fewer interruptions and more relevant suggestions—highlights the value of a lightweight behavioral gate before triggering expensive inference.

\subsubsection{Qualitative Impact on Latency and Attention}

Suppressing unhelpful suggestions carries three downstream advantages:

\begin{itemize}
	\item \textbf{Reduced GPU and network load:} At scale, a 35\% reduction in calls translates directly to lower compute and hosting costs without any model modifications.
	\item \textbf{Fewer unwanted interruptions:} Developers reported fewer unsolicited suggestions during deep-focus tasks, aligning delivery with natural pauses and increasing perceived relevance.
\end{itemize}

Together, these results demonstrate that behavior-driven pre-filtering can both streamline system resource use and respect developer attention, all without touching the LLM itself.

\subsubsection{Behavioral Filtering in Action.}
\begin{figure}[ht]
	\centering
	\includegraphics[width=0.48\textwidth]{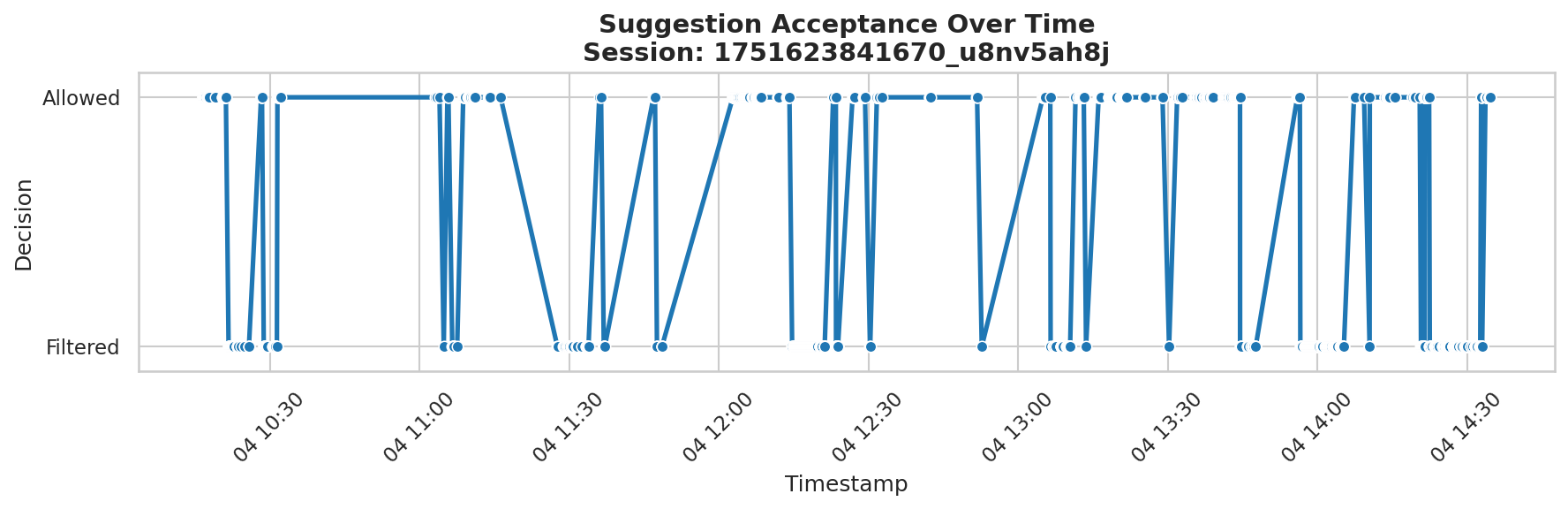}
	\caption{Suggestion filtering decisions over the course of a real developer session. The model dynamically alternates between allowing and suppressing requests in response to real-time behavior. Extended filtering often coincides with focused activity, while suggestions tend to be allowed during pauses or fragmented interaction.}
	\label{fig:session_timeline}
\end{figure}

Figure~\ref{fig:session_timeline} illustrates the gatekeeping behavior of the model during a real developer session. Rather than applying a fixed or periodic policy, the system responds dynamically to interaction context. Suggestions are allowed during slower or exploratory phases, and filtered when telemetry indicates high focus or cognitive load. This adaptivity demonstrates the model's alignment with developer rhythms and its capacity to reduce interruptions without sacrificing useful completions.

\subsection{Statistical Significance of Acceptance‑Rate Gains}
\label{sec:stats}

To verify that the observed jump in suggestion–acceptance rates (from 18.4\% to 34.2\%; cf.\ Section~\ref{sec:acceptancerate}) was not due to sampling variability, we subjected the before–after counts to standard proportion tests. Confidence intervals were computed with the Wilson score method,
and both a pooled two‑proportion \emph{z}-test and an exact
Fisher test were applied at $\alpha=0.05$.

\begin{table}[ht]
	\centering
	\caption{Suggestion acceptance rates before and after behavioural filtering, with 95\% Wilson confidence intervals}
	\label{tab:acceptance-ci}
	\begin{tabular}{lccc}
		\toprule
		\textbf{Period} & \textbf{Accepted / Total} & \textbf{Rate (\%)} & \textbf{95\% CI (\%)} \\
		\midrule
		Before (Baseline)  & 427 / 2\,319 & 18.4 & [16.9, 20.0] \\
		After (Filtered)   & 486 / 1\,422 & 34.2 & [31.8, 36.7] \\
		\bottomrule
	\end{tabular}
\end{table}

\begin{table}[ht]
	\centering
	\caption{Comparative statistics for the two independent proportions}
	\label{tab:stat-tests}
	\begin{tabular}{lc}
		\toprule
		\textbf{Metric} & \textbf{Value} \\
		\midrule
		Absolute difference ($\Delta$)        & +15.8 percentage points \\
		Risk ratio ($p_{\mathrm{after}}/p_{\mathrm{before}}$) & 1.86 \\
		Odds ratio                            & 2.30 \\
		Two‑proportion $z$ statistic          & 10.90 \\
		$p$‑value ($z$‑test)                  & $< 1\times10^{-26}$ \\
		$p$‑value (Fisher exact)              & $< 5\times10^{-26}$ \\
		\bottomrule
	\end{tabular}
\end{table}

\textbf{Interpretation.}  
The large, highly significant $z$ statistic ($|z| = 10.9$, $p \ll 0.001$) and corroborating Fisher exact test confirm that the acceptance‑rate improvement is statistically reliable.  An odds ratio of 2.30 indicates that developers were more than twice as likely to accept a suggestion once behavioral filtering was enabled. The tight, non‑overlapping Wilson intervals further underscore the practical importance of the effect.
These results strengthen our claim that real‑time telemetry can be leveraged not only to cut compute costs but also to deliver meaningfully better assistance to developers.

\subsection{Limitations and Threats to Validity}
\label{sec:limitations}

While the results demonstrate a clear benefit from behavioral pre-filtering, several limitations and threats to validity must be acknowledged.

\paragraph{Participant Pool and Usage Patterns.} The plugin was deployed in an academic research environment with a relatively small pool of users (n=25), of whom only 9 contributed sustained usage data. Developer engagement was voluntary and often intermittent, leading to uneven distribution of suggestion events. As a result, the data may not fully represent the diversity of workflows seen in large-scale industrial settings. Future work will extend deployment to larger and more diverse industrial cohorts to further validate generalization. 

\paragraph{Non-Randomized Deployment.} The evaluation design compares two time periods—before and after model deployment—without random assignment or counterbalancing. Although the conditions were held constant as much as possible, external factors such as developer experience, task type, or familiarity with the plugin may have influenced outcomes. This limits causal claims and motivates future controlled studies.

\paragraph{Behavioral Feature Scope.} Our model relies exclusively on observable interaction data from the VS Code API. While effective, these features do not capture deeper contextual factors such as semantic intent, code quality, or longer-term developer goals. Consequently, the behavioral signal may fail to distinguish between different types of engagement that lead to similar telemetry patterns. In future, richer contextual metrics could complement behavioral telemetry while preserving privacy.

\paragraph{Threshold Sensitivity.} The filtering mechanism depends on a tuned probability threshold to gate LLM requests. Although chosen to balance recall and resource savings, this value may require adjustment in other deployment contexts or user populations. Furthermore, fixed thresholds may not generalize well to developers with distinct working styles. Adaptive or personalized thresholds, potentially learned online, represent a promising path to make filtering more robust across diverse developer styles.

We also note that the deployed model uses a single model trained on pooled data from all participants. While this enables robustness across developers, it does not yet adapt to individual developer profiles. In practice, personalization could be maintained by lightweight per-user models and dynamically fine-tuning thresholds as acceptance/rejection feedback accumulates. We consider this as a promising future direction, and a complement to our current focus on population-level generalization.

Despite these limitations, the system achieved meaningful improvements under realistic conditions, supporting the value of behavioral telemetry as a practical signal for adaptive LLM invocation. Future work can address these concerns by scaling deployment, incorporating user feedback, and exploring more personalized filtering strategies.

\section{Conclusion}
\label{sec:conclusion}

This work addressed an often-overlooked question in LLM-powered programming assistance: \textit{when} to provide a suggestion. We introduced a lightweight, behavioral-only filtering mechanism that predicts the likelihood of code suggestion acceptance from real-time developer telemetry, without inspecting prompt or code content. Integrated into a production-grade Visual Studio Code plugin, the approach achieved two key outcomes in naturalistic use: (i) the suggestion acceptance rate increased from 18.4\% to 34.2\%, and (ii) over 35\% of low-value LLM calls were suppressed. These gains were achieved without any modification to the LLM itself, demonstrating that optimizing \emph{timing} can substantially improve both system efficiency and user experience.

\subsection*{Future Work}
Building on this foundation, several research and engineering directions are promising:

\begin{itemize}
	\item \textbf{Further Personalization and Adaptivity.} Develop per-user or per-session adaptive thresholds, potentially using online learning from ongoing acceptance/rejection feedback to refine responsiveness.
	
    \item \textbf{Richer Models of Developer State.} Extend beyond surface-level interaction metrics to capture higher-order temporal dynamics and transitions between states of focus, exploration, and problem-solving. This could enable more nuanced timing policies that adapt to the ebb and flow of cognitive engagement.

	\item \textbf{Behavioral-Driven Next Edit Prediction.} Build on this pre-filtering framework to anticipate the developer’s likely next editing action, similar to Chen \emph{et al.}’s adaptive next-edit suggestion approach~\cite{chen2025efficientadaptiveeditsuggestion}. Such capability could allow the system not only to time suggestions effectively, but also to shape their content in alignment with imminent editing goals—while remaining within a telemetry-only, privacy-preserving scope.

	\item \textbf{Longitudinal Analysis.} Explore multi-session behavioral trends and persistent user profiles to inform more temporally-aware filtering policies.
\end{itemize}

\subsection*{Closing Remarks}
Our findings highlight that meaningful improvements to AI-assisted coding workflows can be achieved not only by enhancing model quality or prompt design, but also by strategically deciding \emph{when} to assist. By treating developer attention as a first-class resource, behavioral pre-filtering offers a scalable, privacy-preserving, and model-agnostic pathway to reducing wasted computation, improving acceptance rates, and aligning assistance with real-time human context.

\section*{Acknowledgment}
The research was supported by The Russian Science Foundation, agreement №24-11-00272, https://rscf.ru/project/24-11-00272/.

\bibliographystyle{IEEEtran}
\bibliography{references}

\end{document}